\documentclass[
showpacs,preprintnumbers,amsmath,amssymb]{revtex4}

\newcommand{\pslash}{\not \! p}

\newcommand{\kslash}{\not \! k}

\begin{document}

\begin{center}
{\Large{\bf Neutrino-antineutrino Mass Splitting in the Standard Model: Neutrino Oscillation and Baryogenesis
}}
\end{center}
\vskip .5 truecm
\begin{center}
{\bf { Kazuo Fujikawa$^\dagger$ and Anca Tureanu$^*$}}
\end{center}

\begin{center}
\vspace*{0.4cm} {\it {$^\dagger$ Quantum Hadron Physics Laboratory, RIKEN Nishina Center,\\
Wako 351-0198, Japan\\
 $^*$Department of Physics, University of
Helsinki, P.O.Box 64, \\FIN-00014 Helsinki,
Finland\\
}}
\end{center}

\begin{abstract}
By adding a neutrino mass term to the Standard Model, which is Lorentz and $SU(2)\times U(1)$ invariant but non-local to evade $CPT$ theorem,  it is shown that non-locality within a distance scale of the Planck length, that may not be fatal to unitarity in generic effective theory, can generate the neutrino-antineutrino mass splitting of the order of observed neutrino mass differences, which is tested in oscillation experiments, and non-negligible baryon asymmetry depending on the estimate of sphaleron dynamics. The one-loop order induced electron-positron mass splitting in the Standard Model is shown to be finite and estimated  at $\sim 10^{-20}$ eV, well below the experimental bound $< 10^{-2}$ eV. The induced $CPT$ violation in the $K$-meson in the Standard Model is expected to be even smaller and well below the experimental bound $|m_{K}-m_{\bar{K}}|<0.44\times 10^{-18}$ GeV. 

\end{abstract}

\maketitle

\large
\section{CPT theorem and its possible evasion}
The CPT theorem formulated by W. Pauli and G. L\"{u}ders~\cite{pauli}, which is valid for any Lorentz invariant and local theory described by a hermitian Lagrangean with normal spin-statistics, implies the equality of the masses of the particle and antiparticle. Nevertheless, the possible breaking of CPT theorem
has been discussed by many people in the past. To evade CPT theorem, one may consider, for example, \\
1. Non-local theory,\\
2. Lorentz non-invariant theory.\\

We have recently discussed the possible mass splitting of the neutrino and antineutrino in the Standard Model and its physical implications on the basis of Lorentz invariant non-local theory~\cite{fuji-tureanu}. This is a sequel to the analyses of fermion-antifermion mass splitting in a Lorentz invariant non-local theory~\cite{ CFT1, CFT2}. This Lorentz invariant non-local scheme of CPT breaking itself has been revived by the authors in Ref.~\cite{CDNT}; the model considered by them is based on the T-breaking  
with preserved CP, and thus no particle-antiparticle mass splitting. It is sometimes stated in the literature that CPT breaking implies particle-antiparticle mass splitting, but it is not the case; CPT breaking is a necessary condition but not sufficient to generate the particle-antiparticle mass splitting.
The conceptual aspect of this Lorentz invariant non-local scheme has also been clarified in~\cite{JGB}, since it is sometimes {\em erroneously} claimed in the literature that CPT breaking inevitably implies Lorentz symmetry breaking. It was emphasized in \cite{JGB} that the Lorentz invariant $CPT$ breaking scheme, which we adopt in the present study, is a very natural logical possibility. 

The neutrino mass is outside the conventional Standard Model and thus may provide a  window to "brave New World". It may be interesting to incorporate CPT breaking in 
the neutrino mass sector of a minimal extension of the Standard Model.
We incorporate,\\
a)C, CP and CPT breaking,\\
b)Lorentz invariance,\\
c)SU(2)xU(1) gauge invariance,\\
d)Non-locality within a distance scale of the Planck length,\\  
in our model of the neutrino-antineutrino mass splitting~\cite{fuji-tureanu}.

We have shown that sizable neutrino-antineutrino mass splitting, which is readily tested by oscillation experiment, is possible,
but the induced electron-positron mass splitting, for example, is negligibly small and in this sense our scheme of CPT breaking is consistent with existing experimental data~\cite{particle-data}.

\section{ The Model}
 
The Standard Model Lagrangian relevant to our discussion of the electron sector is given by
\begin{eqnarray}\label{standard}
{\cal L}&=&i\overline{\psi}_{L}\gamma^{\mu}
\left(\partial_{\mu} - igT^{a}W_{\mu}^{a}
             - i\frac{1}{2}g^{\prime}Y_{L}B_{\mu}\right)\psi_{L}
\nonumber\\
         & +&i\overline{e}_{R}\gamma^{\mu}(\partial_{\mu}
             + ig^{\prime}B_{\mu})e_{R}
+i\overline{\nu}_{R}\gamma^{\mu}\partial_{\mu}\nu_{R}\nonumber
\\
            &+&[ -
\frac{\sqrt{2}m_{e}}{v}\overline{e}_{R}\phi^{\dagger}\psi_{L}
-\frac{\sqrt{2}m_{D}}{v}\overline{\nu}_{R}\phi_{c}^{\dagger}\psi_{L} -\frac{m_{R}}{2}\nu_{R}^{T}C\nu_{R} + h.c.],
\end{eqnarray}
with {\em assumed} right-handed component $\nu_{R}$. We denote the 
Higgs doublet and its $SU(2)$ conjugate by $\phi$  and $\phi_{c}\equiv i\tau_{2}\phi^{\star}$, respectively.
We tentatively set $m_{R}=0$ with enhanced lepton number symmetry, namely, a "Dirac neutrino". In short, we assume that every mass arises from the Higgs doublet, which has been discovered recently.

One may  add a hermitian {\bf non-local} Higgs coupling with a real parameter $\mu$ to the above Lagrangian~\cite{fuji-tureanu},
\begin{eqnarray}\label{(5)}
{\cal L}_{CPT}(x)
&&=-i\frac{2\sqrt{2}\mu}{v}\int
d^{4}y\Delta_{l}(x-y)\theta(x^{0}-y^{0})\nonumber\\
&&\times\{\bar{\nu}_{R}(x)\left(\phi_{c}^{\dagger}(y)\psi_{L}(y)\right)
-\left(\bar{\psi}_{L}(y)\phi_{c}(y)\right)\nu_{R}(x)\},
\end{eqnarray}
without spoiling Lorentz invariance and $SU(2)_{L}\times U(1)$ gauge symmetry. 
 Here we defined a "time-like non-local factor",
\begin{eqnarray}\label{(6)}
\Delta_{l}(x-y)\equiv \delta\left((x-y)^{2}-l^{2}\right)-\delta\left((x-y)^{2}-{l^{\prime}}^{2}\right)
\end{eqnarray}
with $l$ standing for fixed length scale and $l^{\prime}=0$, for simplicity.

In the unitary gauge, the neutrino mass term 
becomes
\begin{eqnarray}\label{mass}
S_{\nu \rm mass}
&=&\int d^{4}x\Big\{-m_{D}\bar{\nu}(x)\nu(x)\left(1+\frac{\varphi(x)}{v}\right)\nonumber\\
&& -i\mu\int
d^{4}y\Delta_{l}(x-y)\left[\theta(x^{0}-y^{0})-\theta(y^{0}-x^{0})\right]\bar{\nu}(x)\nu(y)\nonumber\\
&&+i\mu\int
d^{4}y\Delta_{l}(x-y)\bar{\nu}(x)\gamma_{5}\nu(y)\nonumber
\\
&&-i\frac{\mu}{v}\int
d^{4}y\Delta_{l}(x-y)\theta(x^{0}-y^{0})\nonumber\\
&&\times\left[\bar{\nu}(x)(1-\gamma_{5})\nu(y)-\bar{\nu}(y)(1+\gamma_{5})\nu(x)\right]\varphi(y)\Big\}. 
\end{eqnarray}
The term 
\begin{eqnarray}\label{(8)}
&&-i\mu\int
d^{4}x\int d^{4}y\Delta_{l}(x-y)\left[\theta(x^{0}-y^{0})-\theta(y^{0}-x^{0})\right]
\bar{\nu}(x)\nu(y)
\end{eqnarray}
in the action  preserves $T$ but has $C=CP=CPT=-1$ and thus gives rise to particle-antiparticle mass splitting.

The equation of motion for the free neutrino is given by 
\begin{eqnarray}\label{(9)}
i\gamma^{\mu}\partial_{\mu}\nu(x)
&=&m_{D}\nu(x)\nonumber\\
&+&i\mu\int
d^{4}y\Delta_{l}(x-y)\left[\theta(x^{0}-y^{0})-\theta(y^{0}-x^{0})\right]\nu(y)\nonumber\\
&-&i\mu\int d^{4}y\Delta_{l}(x-y)\gamma_{5}\nu(y).
\end{eqnarray}

By inserting an Ansatz,
$\nu(x)=e^{-ipx}U(p)$,
we obtain
\begin{eqnarray}\label{(2.6)}
\pslash U(p)&=&\Big\{m
+i[f_{+}(p)-f_{-}(p)]-i g(p^{2})\gamma_{5}\Big\}U(p),
\end{eqnarray}
where 
\begin{eqnarray}\label{(1.3)}
&&f_{\pm}(p)=\mu\int d^{4}z
e^{\pm ipz}\theta(z^{0})\left[\delta\left((z)^{2}-l^{2}\right)-\delta(z^{2})\right],\nonumber\\
&&g(p^{2})=
\mu\int d^{4}ze^{ipz}[\delta((z)^{2}-l^{2})-\delta((z)^{2})].
\end{eqnarray}
The last term is  parity violating mass term, which is $C$ and $CPT$ preserving.

The factor $f_{\pm}(p)$ is mathematically related to
the two-point Wightman function,
\begin{eqnarray}\label{(13)}
\langle 0|\phi(x)\phi(y)|0\rangle=\int d^{4}p
e^{i(x-y)p}\theta(p^{0})\delta(p^{2}-m^{2}).
\end{eqnarray}
We know the properties of the Wightman function well, and they are useful in our analysis. For example, the Wightman function has a quadratic divergence for the short distance, which is independent of mass. This implies that 
our CPT breaking term in the Dirac equation is free of quadratic divergence in the infrared.

For  time-like $p^{2}>0$, one may go to the frame where $\vec{p}=0$,
\begin{eqnarray}\label{(14)}
p_{0}&=&\gamma^{0}[m_{D}-f(p_{0})-ig(p^{2}_{0})\gamma_{5}],
\end{eqnarray}
with
\begin{eqnarray}\label{(15)}
f(p_{0})
&\equiv&-i[f_{+}(p_{0})-f_{-}(p_{0})]\nonumber\\
&=&4\mu\pi\int_{0}^{\infty}dz\Big\{\frac{z^{2}\sin [p_{0}\sqrt{z^{2}+l^{2}}]}{\sqrt{z^{2}+l^{2}}}-\frac{z^{2}\sin [p_{0}\sqrt{z^{2}}]}{\sqrt{z^{2}}}\Big\},\nonumber\\
g(p^{2}_{0})&=&4\mu\pi\int_{0}^{\infty}dz\Big\{\frac{z^{2}\cos [p_{0}\sqrt{z^{2}+l^{2}}]}{\sqrt{z^{2}+l^{2}}}-\frac{z^{2}\cos [p_{0}\sqrt{z^{2}}]}{\sqrt{z^{2}}}\Big\}.
\end{eqnarray}
For space-like $p^{2}<0$, one can confirm that the $CPT$ violating term vanishes, $f(p)=0$, by choosing $p_{\mu}=(0, \vec{p})$.\\

Since we are assuming that the $CPT$ breaking terms are small, we may
solve the mass eigenvalue equations in (10) iteratively
\begin{eqnarray}\label{(17)}
m_{\pm}
&\simeq&m_{D}-i\gamma_{5} g(m_{D}^{2}) \pm
f(m_{D}).
\end{eqnarray}
The parity violating mass $-i\gamma_{5} g(m_{D}^{2})$ is now transformed away by a suitable global chiral transformation. 
In this way, the neutrino-antineutrino mass splitting is incorporated in the
Standard Model by the Lorentz invariant non-local $CPT$ breaking
mechanism, without spoiling the $SU(2)_{L}\times U(1)$ gauge symmetry. 

The
Higgs particle $\varphi$ itself has a tiny $C$, $CP$ and $CPT$
violating coupling.

\section{ Evaluation of mass splitting}
The CPT violating term is evaluated as 
\begin{eqnarray}\label{(20)}
f(p)
&=&-4\pi\mu l^{2}[\theta(p_{0})-\theta(-p_{0})]\theta(p^{2})\nonumber\\
&&\times\Big\{\int_{1}^{\infty}du \frac{1}{2u(\sqrt{u^{2}-1}+u)^{2}}\sin (|p_{0}|l u)\nonumber\\
&&-\frac{1}{2}\int_{0}^{1}du \frac{\sin (|p_{0}|l u)}{u} +\int_{0}^{1}du u
\sin (|p_{0}|lu)+\frac{1}{2}\int_{0}^{\infty}du \frac{\sin (u)}{u}\Big\}.
\end{eqnarray}
Our $CPT$ violating term is characterized by the quantity  
\begin{eqnarray}\label{(21)}
\mu l^{2},
\end{eqnarray}
which has the dimension of mass.

For $|p_{0}|l\ll 1$, which is generally expected since we are going to choose $l$ at about the Planck length, we have Lorentz invariant,
\begin{eqnarray}\label{(22)}
f(p)
&\simeq&-\pi^{2}\mu l^{2}[\theta(p_{0})-\theta(-p_{0})]\theta(p^{2}).
\end{eqnarray}
Thus the neutrino-antineutrino mass splitting is given by 
\begin{eqnarray}\label{(23)}
\Delta m &\simeq&2\pi^{2}\mu l^{2}.
\end{eqnarray}
Our $CPT$ violating term $f(p_{0})$ is odd in $p_{0}$ and 
$f(\pm 0)=\mp \Delta m/2$. 

As for the parity violating mass term, we have (in the frame with $\vec{p}=0$ for $p^{2}>0$)
\begin{eqnarray}\label{(25)}
g(p^{2})
&=&-4\pi\mu l^{2}
\Big\{\int_{1}^{\infty}du\frac{1}{\sqrt{u^{2}-1}+u}\cos [p_{0}lu]\nonumber\\
&& +\frac{\sin [p_{0}l]}{p_{0}l}+\frac{\cos [p_{0}l]-1}{(p_{0}l)^{2}}\Big\}.
\end{eqnarray}
This formula is again well-defined if precise $p_{0}=0$ is excluded.

\section{Neutrino-antineutrino mass splitting}

The Lorentz invariant non-local factor   
\begin{eqnarray}\label{(27)}
\left[\theta(x^{0}-y^{0})-\theta(y^{0}-x^{0})\right]
\left[\delta((x-y)^{2}-l^{2})-\delta((x-y)^{2})\right],
\end{eqnarray}
mostly cancels out the infinite time-like volume effect and eliminates the quadratic
infrared divergence completely. In effect, non-locality is limited within the fluctuation around the tip of the light-cone characterized by the length scale $l$, which we choose to be the Planck length.

By setting 
\begin{eqnarray}
l=1/M_{P},\ \ \  \mu=M^{3},
\end{eqnarray}
 the neutrino-antineutrino mass splitting is given by 
\begin{eqnarray}\label{(28)}
\Delta m = 2\pi^{2} \mu l^{2}= 2\pi^{2}M(M/M_{P})^{2},
\end{eqnarray}
which may be regarded as a gravitational effect due to the Newton constant $G_{N}=1/M_{P}^{2}$.
If one chooses $M\sim 10^{9}$ GeV, the neutrino-antineutrino  mass splitting becomes of the order of the observed neutrino mass (difference) $\sim 0.1$ eV.
The possible neutrino-antineutrino  mass splitting has been discussed in the past in connection with neutrino oscillation phenomenology~\cite{pakvasa, murayama, barenboim}.
The neutrino-antineutrino mass splitting 
\begin{eqnarray}
\Delta m=10^{-1}\sim 10^{-2} eV,
\end{eqnarray}
which is intended to be  of the order of $m_{D}/5$, is generated by $M\simeq 10^{8}\sim 10^{9}$ GeV and  appears to be allowed by the presently available experimental data such as MINOS~\cite{minos1}. \\

\noindent{\bf Baryogenesis}\\

A neutrino-antineutrino mass difference
would result in a leptonic  matter-antimatter asymmetry proportional to the mass difference. This asymmetry is transmitted to the baryon sector through the chiral anomaly and  sphaleron processes which preserve $B - L$ but violate $B + L$. 

This "kinematical" picture implies the asymmetry
in the neutrino and antineutrino of the order~\cite{zeldovich}
\begin{eqnarray}
(n_{\nu}-n_{\bar{\nu}})/ n_{\nu}\simeq m_{D}\Delta m/T^{2},
\end{eqnarray}
which is, however, too small at the electroweak energy scale to generate the baryon asymmetry via sphaleron processes in our case with  $\Delta m=10^{-1}\sim 10^{-2}$ eV. 
Besides, this initial asymmetry requires the lepton number non-conservation, while the lepton number is conserved in our model without sphaleron effects.

Thus the lepton and quark sectors need to be treated simultaneously in the presence of sphalerons. Barenboim, Borissov, Lykken and Smirnov discuss a rather elaborate sphaleron dynamics and conclude at weak scale $M_{W}$~\cite{barenboim}
\begin{eqnarray}\label{(30)}
\frac{n_{B}}{n_{\gamma}}\sim \frac{\Delta m}{M_{W}}.
\end{eqnarray}
This estimate in the present case with $\Delta m=10^{-1}\sim 10^{-2}$ eV, namely $n_{B}/n_{\gamma}\sim 10^{-12} - 10^{-13}$, is smaller than the observed value $n_{B}/n_{\gamma} \simeq 10^{-10}$, but it still gives an interesting number.

This {\em  equilibrium electroweak baryogenesis} does not need 
CP violation other than for the purpose of producing neutrino-antineutrino mass splitting. This mechanism differs from the more conventional baryogenesis ~\cite{sakharov, yoshimura} or leptogenesis~\cite{fukugita}.

\section{ Higher order effects}

The propagator of the neutrino in path integral, which is based on Schwinger's action principle~\cite{fujikawa}, is given by~\cite{CFT1}, 
\begin{eqnarray}\label{(31)}
\langle T^{\star}\nu(x)\bar{\nu}(y)\rangle
&=&\int \frac{d^{4}p}{(2\pi)^{4}} e^{-ip(x-y)}\nonumber\\
&&\times\frac{i}{\pslash-m_{D} +i\epsilon+i\gamma_{5}g(p^{2})+f(p)}.
\end{eqnarray}
We can show 
\begin{eqnarray}
&&f(p)=-i[f_{+}(p)-f_{-}(p)]\rightarrow 0,\\
&&g(p^{2}) \rightarrow 0
\end{eqnarray}
for $p\rightarrow\infty$ in Minkowski space, which is an analogue of the cluster property of the Wightman function in (9).
The propagator for Minkowski momentum is thus well behaved and the effects of non-locality are mild and limited. One may thus be tempted to replace $T^{\star}$ product by the canonical $T$ product~\cite{fujikawa} in (24).

In the analysis of the renormalization, however, it is customary to consider the Euclidean amplitude obtained from the Minkowski amplitude by Wick rotation. Our propagator, which contains trigonometric functions, has undesirable behavior under the Wick rotation such as 
\begin{eqnarray}
\sin p_{0}z\rightarrow i\sinh p_{4}z,
\end{eqnarray}
and exponentially divergent behavior is generally induced and the effects of non-locality become significant. 

One might still argue that higher order effects in field theory defined on Minkowski space are in principle analyzed in Minkowski space and, if that is the case, our propagator suggests the ordinary renormalizable behavior.
This issue is left for the future study.\\

\noindent{\bf Induced electron-positron mass splitting}
\\

Our Lorentz invariant $CPT$ violating term is effectively replaced by
\begin{eqnarray}\label{(33)}
f(p)=-\pi^{2}\mu l^{2}[\theta(p_{0})-\theta(-p_{0})]\theta(p^{2}),
\end{eqnarray}
which is similar to a constant mass term except for the  $CPT$ violating 
factor $[\theta(p_{0})-\theta(-p_{0})] \theta(p^{2})$.
When this term is inserted into a neutrino line in
Feynman diagrams of the Standard Model, those Feynman diagrams are expected to show ordinary high energy behavior for a mass insertion, if the naive power counting works. Also, perturbative unitarity may not be spoiled since $f(p)$ is essentially constant in momentum space. 

We examine the electron-positron mass splitting induced by the above factor $f(p)$, when inserted into a neutrino line in one-loop self-energy diagrams of the electron in the Standard Model.
The $W$-boson contribution is then given by   
\begin{eqnarray}\label{(35)}
&&g^{2}\int\frac{d^{4}p}{(2\pi)^{4}}[\gamma^{\alpha}\frac{(1-\gamma_{5})}{2}
\frac{\pslash+m_{D}}{p^{2}-m^{2}_{D} +i\epsilon}f(p)\frac{\pslash+m_{D}}{p^{2}-m^{2}_{D} +i\epsilon}\gamma_{\alpha}\frac{(1-\gamma_{5})}{2}]\nonumber\\
&&\hspace{2cm}\times\frac{1}{(k-p)^{2}-M_{W}^{2} +i\epsilon}.
\end{eqnarray}
 We thus obtain a  finite result, $$\sim\alpha [m_{D}\kslash/M_{W}^{2}](\mu l^{2})[\theta(k^{0})-\theta(-k^{0})] \theta(k^{2})$$ with $\alpha$ standing for the fine structure constant, which in fact gives the leading contribution.

The induced $CPT$ violating effect on the electron- positron splitting is {\bf finite} and it is estimated at the order\cite{fuji-tureanu},
\begin{eqnarray}\label{(39)}
\alpha [m_{D}m_{e}/M_{W}^{2}](\mu l^{2})[\theta(k^{0})-\theta(-k^{0})] \theta(k^{2}),
\end{eqnarray}
which, for $\pi^{2}\mu l^{2}=10^{-1}\sim 10^{-2}$ eV, is 
\begin{eqnarray}
|m_{e}-m_{\bar{e}}|\sim 10^{-20} eV,
\end{eqnarray}
and thus well below the present experimental upper bound $\leq 10^{-2}$ eV~\cite{particle-data}.

The induced $CPT$ violation is expected to be smaller in the quark sector (as a two-loop effect) than in the charged leptons in the $SU(2)\times U(1)$ invariant theory, and thus much smaller than the well-known limit on the $K$-meson~\cite{particle-data}, 
\begin{eqnarray}
|m_{K}-m_{\bar{K}}|<0.44\times 10^{-18} GeV.
\end{eqnarray}

\section{Conclusion}

Our proposed model of Lorentz invariant non-local CPT breaking allows sizable neutrino-antineutrino mass splitting, which can be {\em tested by oscillation experiments},  without inducing detectable undesirable effects in other sectors of the Standard Model. Also, it has a potentially interesting implication on baryogenesis.
The Lorentz invariant non-local CPT breaking at the Planck scale thus suggests a promising 
CPT breaking scheme for an effective field theory, although a deeper analysis of basic issues such as unitarity is required. The origin of CPT breaking at the Planck scale itself remains to be clarified. 
\\
\\
{\em Note added}: A. Suzuki of KamLAND and A.Y. Smirnov informed us that the sun neutrino data and the KamLAND antineutrino data show a discrepancy in the neutrino mass of $\sim e^{-3} eV$ which is a $2\sigma$ effect. It is an interesting subject to examine this and other related neutrino oscillation experiments in connection with the test of CPT symmetry. We thank them for this interesting information.

\end{document}